\def\dash{\hbox{--}}
\def\ts{\thinspace}           

\def\puncspace{\ifmmode\,\else{\ifcat.\C{\if.\C\else\if,\C\else\if?\C\else%
\if:\C\else\if;\C\else\if-\C\else\if)\C\else\if/\C\else\if]\C\else\if'\C%
\else\space\fi\fi\fi\fi\fi\fi\fi\fi\fi\fi}%
\else\if\empty\C\else\if\space\C\else\space\fi\fi\fi}\fi}
\def\SP{\let\\=\empty\futurelet\C\puncspace}

\def\ee#1{\ifmmode {} \times 10^{#1} \else ${} \times 10^{#1}$\fi}
\def\sub#1{\ifmmode _{#1} \else $_{#1}$\fi}
\def\sup#1{\ifmmode ^{#1} \else $^{#1}$\fi}

 \mathcode`*="002A   
\def\simless{\mathbin{\lower 3pt\hbox
     {$\rlap{\raise 5pt\hbox{$\char'074$}}\mathchar"7218$}}}   
\def\simmore{\mathbin{\lower 3pt\hbox
     {$\rlap{\raise 5pt\hbox{$\char'076$}}\mathchar"7218$}}}   

\def\about{$\sim$\thinspace}
\def\aboutless{$\simless$\thinspace}

\def\etal{{et al.}\SP}

\def\keV{{\hbox{keV}}\SP}

\def\pcc{{\hbox{cm\sup{-3}}}\SP}

\def\psqcm{{\hbox{cm\sup{-2}}}\SP}

\def\cyg#1{\leavevmode\hbox{Cyg~X-#1}\SP}
\def\gx#1{\leavevmode\hbox{GX~#1}\SP}

\def\sco#1{\leavevmode\hbox{Sco~X-#1}\SP}

\def\exosat{\leavevmode{EXOSAT}\SP}

\def\e*{{\ifmmode E^* \else {$E^*$}\fi}}
\def\emax{{\ifmmode E_{\rm max} \else {$E_{\rm max}$}\fi}}
\def\le{{\ifmmode L_E \else {$L_E$}\fi}}
\def\lsun{{\ifmmode L_\odot \else {$L_{\odot}$}\fi}}
\def\mdot{{\ifmmode \dot M \else {$\dot M$}\fi}}
\def\mdote{{\ifmmode \dot M_E \else {$\dot M_E$}\fi}}
\def\msun{{\ifmmode M_\odot \else {$M_{\odot}$}\fi}}
\def\sun{{\ifmmode _\odot \else {$_{\odot}$}\fi}}
\def\tauhcc{{\ifmmode \tau_{es}^{\scriptscriptstyle HCC}
            \else {$\tau_{es}^{\scriptscriptstyle HCC}$}\fi}}
\def\taurf{{\ifmmode \tau_{es}^{\scriptscriptstyle RF}
           \else {$\tau_{es}^{\scriptscriptstyle RF}$}\fi}}
\def\tehcc{{\ifmmode T_e^{\scriptscriptstyle HCC}
          \else {$T_e^{\scriptscriptstyle HCC}$}\fi}}
\def\yhcc{{\ifmmode y^{\scriptscriptstyle HCC}
          \else {$y^{\scriptscriptstyle HCC}$}\fi}}

\documentstyle[aas2pp4]{article}

\slugcomment{TEXT}

\lefthead{Psaltis, Lamb, \& Miller}
\righthead{X-Ray Spectra of Z Sources}

\begin{document}

\author{Submitted to {\em The Astrophysical Journal (Letters)}, 1995
May 30}
\vspace*{1.5cm}

\title{X-Ray Spectra of Z Sources}
\vspace*{.5cm}

\author{Dimitrios Psaltis\altaffilmark{1} and
        Frederick K. Lamb\altaffilmark{1}}
\affil{Physics Department,
       University of Illinois at Urbana-Champaign,\\
      1110 W. Green St., Urbana, IL 61801, USA\\
      demetris@astro.uiuc.edu and f-lamb@uiuc.edu}

\vspace*{0.3cm}
\and
\vspace*{-0.3cm}

\author{Guy S. Miller}
\affil{Department of Physics and Astronomy, Northwestern University,\\
       2145 Sheridan Road, Evanston, IL 60208, USA\\
       gsmiller@casbah.acns.nwu.edu}

\altaffiltext{1}{Also, Astronomy Department, University of Illinois at
       Urbana-Champaign}

\vspace*{0.3cm}
\begin{abstract}
A simple, physically consistent model has been proposed that seeks to
explain in a unified way the X-ray spectra and rapid X-ray
variability of the so-called Z sources and other accreting neutron
stars in low-mass systems. Here we summarize the results of detailed
numerical calculations of the X-ray spectra of the Z sources
predicted by this model. Our computations show that in the Z sources,
photons are produced primarily by electron cyclotron emission in the
neutron star magnetosphere. Comptonization of these photons by the
hot central corona and radial inflow produces X-ray spectra,
color-color tracks, and countrate variations like those observed in
the Z sources.
\end{abstract}

\keywords{radiation mechanisms: thermal -- stars: neutron -- X-rays:
stars}

\section{INTRODUCTION}

The six known Z sources (Hasinger \& van der Klis 1989) are among the
most luminous low-mass X-ray binaries. They are unique in tracing out
Z-shaped tracks in X-ray color-color diagrams on time scales of
minutes to hours (see van der Klis 1989; van der Klis \& Lamb 1995).
The tracks are thought to be produced by variations in the mass
accretion rate, which appears to increase monotonically as a Z source
moves from its horizontal to its normal and then to its flaring
branch (see van der Klis \& Lamb 1995). Power spectra of Z-source
brightness variations show two types of quasi-periodic oscillations
(QPOs) and three types of aperiodic flickering (``noise''). The
properties of the QPOs and flickering vary systematically with the
position of a source on its Z track.

The X-ray spectra of the Z sources and the origin of their Z-shaped
tracks has been a puzzle. As the precision of X-ray spectral
measurements has improved, a variety of increasingly complicated
mathematical expressions and physical models has been used in
attempts to obtain statistically acceptable fits to the spectra of Z
sources and other accreting neutron stars in low-mass systems (see,
e.g., Swank \& Serlemitsos 1985; Hirano \etal\ 1984; Mitsuda \etal\
1984, 1989; Czerny, Czerny, \& Grindlay 1986; White \etal\ 1986;
Vacca \etal\ 1987; White, Stella, \& Parmar 1988; Ponman, Foster, \&
Ross 1990; Christian \& Swank 1995). No attempt was made in these
works to relate the spectral properties of a source to its rapid
X-ray variability, and many of the models are not physically
consistent (see Lamb 1986; White \etal\ 1988). In these early works,
models were fit to X-ray spectra averaged over hours, and the origin
of the Z tracks of the Z sources was not addressed (in some cases the
existence of the Z tracks was unknown at the time the work was done).

Several authors have shown that the tracks of one or two Z sources
can be reproduced by varying the parameters in various complicated
expressions (Schulz, Hasinger, \& Tr\"umper 1989; Hasinger \etal\
1990; Hoshi \& Mitsuda 1991; Honma \& Mineshige 1993), but the
required variations of the parameters are typically erratic and
not motivated by physical arguments. Schulz \& Wijers (1993)
showed that suitable variations of the parameters in a model which
assumes that blackbody emission from the neutron star surface is
Comptonized by electrons in an extensive hot corona can reproduce
the observed color-color tracks of four of the Z sources. However,
the model is not physically self-consistent and the inferred
variations of the parameters along the Z tracks are erratic and
unrelated to the other properties of the source.

A model that seeks to provide a unified, physically consistent
explanation of the X-ray spectra and rapid X-ray variability of the Z
sources as well as other low-mass binary systems containing accreting
neutron stars has been proposed by Lamb (1989, 1995). In this
`unified model' the Z sources are rapidly spinning neutron stars with
magnetic fields \about 10\sup8--10\sup9~G, accreting matter from
their companions at \about 0.5--1.1 times the Eddington critical rate
\mdote\ via a Keplerian disk and an approximately radial inflow that
originates in a corona above the inner disk. The radial flow supplies
\about 20\% of the total mass flux. Interaction of the accretion disk
and radial flow with the \about 3\ee6~cm neutron star magnetosphere
produces a hot central corona (HCC) around the magnetosphere. Soft
(\about 0.5--0.8~keV) photons produced within the magnetosphere by
high-harmonic cyclotron emission and other processes are Comptonized
by electrons in the HCC and the radial inflow. The interaction of
radiation with the radial flow keeps its temperature close to the
local Compton temperature (\about 1 \keV). The result is a relatively
flat X-ray spectrum that cuts off near the \about 10~keV temperature
of the electrons in the HCC. Preliminary results (see Lamb 1989)
showed that the X-ray spectra and colors predicted by the unified
model are roughly consistent with the observed spectra and colors of
the Z sources.

When the accretion rate is \about 0.5--0.9\ts\mdote, interaction of
the small magnetosphere with the accretion disk produces a
quasi-periodic luminosity oscillation at the beat frequency between
the stellar spin frequency and the orbital frequency of clumps near
the inner edge of the disk (Alpar \& Shaham 1985; Lamb \etal\ 1985;
Shibazaki \& Lamb 1987). This is the so-called horizontal branch
oscillation (HBO). When the mass accretion rate rises above $\sim
0.8\,\mdote$, a soft, global, radiation-hydrodynamic mode of the
radial inflow becomes weakly damped and is therefore excited by
fluctuations in the accretion flow, causing the X-ray spectrum to
oscillate quasi-periodically at a frequency approximately equal to
the inverse of the inflow time from the outer radius of the flow
(Lamb 1989, 1995; Fortner et al. 1989, 1995; Miller \& Lamb 1992). At
still higher accretion rates, nonradial modes develop (Lamb 1989,
1995; Miller \& Park 1995). This is the so-called normal/flaring
branch oscillation (N/FBO).

In this {\em Letter\/} we summarize detailed numerical calculations
of the X-ray spectra of the Z sources based on the unified model. Our
results provide a simple, physically consistent explanation of the
spectra of the Z sources, including their Z tracks, as well as other
previously unexplained properties of these sources. A more detailed
account of the spectral model and further applications to the Z
sources and other types of accreting neutron stars in low-mass
systems will be presented elsewhere.

\section{SPECTRAL MODEL}

We have computed the spectrum of the radiation produced in the
magnetosphere, taking into account resonant and nonresonant
scattering as well as cyclotron absorption and emission (Hartmann
\etal\ 1988) and the effects of induced processes, which typically
dominate spontaneous processes by factors \about 20. We assume that
the intrinsic stellar magnetic field is dipolar and compute the
variation of the magnetic field with radius, treating the
magnetospheric boundary as spherically symmetric. The spectrum of the
radiation that emerges from the magnetosphere depends strongly on the
field strength and electron temperature and more weakly on the
electron density. For stellar magnetic fields \about 10\sup9~G,
electron temperatures \about 5--15~keV, and electron column densities
\about 10\sup{19}--10\sup{20}~\psqcm, the spectrum is roughly
triangular, following the Planck curve until somewhat below $\emax
\sim 0.1 \dash 1~\keV$ and then falling steeply as the thermalization
length becomes larger than the magnetosphere.

Photons emerging from the magnetosphere into the HCC are Comptonized
by the hot electrons there, losing energy to electron recoil and
gaining energy from electron thermal and bulk motion (free-free
absorption and emission are negligible at the densities,
temperatures, and photon energies of interest). The qualitative shapes
of Z-source X-ray spectra (Schulz \etal\ 1989) suggest that \tehcc,
the electron temperature in the HCC, varies from 5 to 15~keV,
depending on the source and its position on its Z track. The shapes
of Z-source X-ray spectra and the low upper limits on the amplitudes
of any periodic X-ray brightness oscillations (Vaughan \etal\ 1994)
require that \tauhcc, the electron scattering optical depth of the
HCC, be \about 5--7 (Brainerd \& Lamb 1987; Lamb 1989, 1995). Because
\tauhcc\ is tightly constrained by these considerations, we set it
equal to 6 in the present calculations.

Since \tehcc\ is much greater than \emax, the spectrum of the
radiation that emerges from the HCC above 1~keV is a slightly rounded
power law that cuts off at \about \tehcc\ (see Rybicki \& Lightman
1979). The exponent of the power law depends primarily on $\yhcc
\equiv (4 k_B\tehcc/m_e c^2) (\tauhcc)^2$, the Compton $y$ parameter
in the HCC. The mean energy of the photons leaving the HCC is a
factor $\sim e^y$ greater than the mean energy of the photons
entering the HCC, so the luminosity of the HCC is $L \approx e^y
L_{\rm m}$, where $L_m$ is the luminosity of the magnetosphere.

Interaction of the radiation from the HCC with the radial flow does
not affect the luminosity (the thermal content of the flow is
negligible and hence the temperature of the electrons in the radial
flow is equal to the local Compton temperature; see Lamb 1989, 1995),
but it does affect slightly the shape of the spectrum, increasing the
curvature of the spectrum at \about 2--5~keV. The qualitative shapes
of Z-source X-ray spectra constrain \taurf, the electron scattering
optical depth of the radial flow, to be in the range 2--10; the
radiation-hydrodynamic model of the N/FBO further constrains \taurf\
to be \about 6 near the middle of the normal branch (Fortner \etal\
1989, 1995; Miller \& Lamb 1992).

In computing the X-ray spectra presented here, we approximated the
spectrum of the radiation coming from the magnetosphere by a
triangular function that mimicks the results of our detailed
numerical computations and calculated the effects of Compton
scattering in the HCC and radial flow using the approach of Miller \&
Lamb (1992), which treats these regions as uniform. The finite size
and spherical geometry of the HCC and radial flow are taken into
acount by specifying appropriate distributions of photon escape
times. The effect of the converging flow was neglected. For simplicity
we assumed that the electrons in the magnetosphere and HCC have the
same temperature. To compare our calculated spectra with spectra
observed with \exosat, we used the interstellar absorption
cross-section of Morrison \& McCammon (1983) and the appropriate
energy response matrix of the \exosat ME detector (see Turner,
Smith, \& Zimmerman 1981).

\section{RESULTS}

The fundamental spectral parameters of the unified model are the mass
$M$, radius $R$, and magnetic field $B$ of the neutron star, and the
mass accretion rate \mdot. The X-ray spectrum observed at Earth also
depends on two additional `external' parameters: the distance $D$
and the neutral hydrogen column density $N_H$ between the Earth and
the source. Of these six parameters, only \mdot\ is expected to vary
significantly on the time scales of interest here.

In practice, some of the physical processes that are important near
the neutron star are fairly well understood whereas others are
not. For this reason we work with the three secondary quantities
that directly enter our computation of the X-ray spectrum observed by
\exosat, namely, the peak energy \emax\ of the spectrum emerging from
the magnetosphere, the Compton parameter \yhcc\ of the HCC, and the
scattering optical depth \taurf\ of the radial flow. As discussed in
\S~2, these parameters are tightly constrained by the qualitative
shape of Z-source X-ray spectra and the rapid X-ray variability of
these sources. Once the three secondary spectral parameters are
determined, approximate physical relations can be used to estimate
the corresponding values of $B$ and \mdot, thereby checking the
consistency of the unified model. We determine the external constant
$N_H$ from observations other than the ones with which we compare our
predicted spectra.

We specify the position of a Z source on its track by a {\em rank
number\/}, defined to be 1.0 at the junction of the horizontal and
normal branches and 2.0 at the junction of the normal and flaring
branches (see van der Klis \& Lamb 1995). Figure~1 shows how \emax,
\yhcc, and \taurf\ are expected to vary with rank number as \mdot\
increases from 0.7\ts\mdote\ to 1.1\ts\mdote, in the unified model
(see below). Figure~2 compares the resulting color-color and
hardness-countrate tracks with \cyg2 tracks and the predicted
countrate spectrum at rank number 1.5 with the countrate spectrum
observed near the middle of the normal branch. The agreement is quite
good, given the simplicity of the model. We now briefly describe the
physical reasons for the changes in the spectral parameters shown in
Figure~1, and explain why these changes produce the Z tracks shown in
Figure~2.

{\em Horizontal branch}.---As \mdot\ increases from \about
0.7\ts\mdote\ to \about 0.9\ts\mdote, the magnetosphere is
compressed, but \emax\ remains nearly constant because it is most
sensitive to the magnetic field at \about 2--3 stellar radii, which
remains almost unchanged. The cyclotron luminosity increases more
slowly than the accretion luminosity, so the electron temperature
rises in order to maintain thermal balance. The resulting increase in
\yhcc\ tends to increase both the soft and hard colors. The increase
of the mass flux in the radial flow causes \taurf\ to increase, which
tends to increase the countrate in the 2--5~keV energy range and
hence to increase the soft color and decrease the hard color. The net
effect is to produce an approximately horizontal track in color-color
and hardness-countrate diagrams.

{\em Normal branch}.---As \mdot\ increases from \about 0.9\ts\mdote\
to \about 1.0\ts\mdote, the magnetosphere is further compressed. The
increase in the mass flux and the rapid reduction of the inflow
velocity caused by the increase in the outward radiation force cause
$n_e$ to increase by a factor \about 8, causing \emax\ to increase by
\about 10\%. As a result, the cyclotron luminosity increases faster
than the accretion luminosity, and hence the electron temperature
must fall in order to maintain thermal balance. The resulting
decrease in \yhcc\ causes the spectrum to become steeper, decreasing
the soft and the hard colors. Meanwhile, \taurf\ continues to rise.
As \yhcc\ decreases, fewer cyclotron photons are scattered up into
the \exosat energy range and hence the countrate decreases, even
though the luminosity is increasing. The result is a diagonal track
running from upper left to lower right in both color-color and
hardness-countrate diagrams.

{\em Flaring branch}.---The character of the accretion flow when
\mdot\ exceeds \mdote\ is uncertain, but is probably aspherical and
time-dependent, with matter flowing inward in some regions and
outward in others (Lamb 1989, 1995). In such a situation, radiation
escapes primarily through the regions of lower optical depth and the
magnetosphere expands as \mdot\ increases, decreasing \emax\ and
increasing \yhcc; \taurf\ continues to increase. The result is a
flaring branch that is nearly parallel to the normal branch but
extends to a slightly greater soft color.

\section{DISCUSSION}

Our results show that the unified model, which predicts that
self-absorbed cyclotron emission in the neutron star magnetosphere is
Comptonized by electrons in a hot central corona and radial inflow,
produces X-ray spectra that agree with those of \cyg2. Variation of
\mdot\ from 0.7\ts\mdote\ to 1.1\ts\mdote\ produces Z-shaped tracks
in X-ray color-color diagrams (CDs) and hardness-countrate diagrams
(HCDs) that agree with the tracks made by \cyg2.

Analysis of archival \exosat\ data (Kuulkers 1995) suggests that the
Z sources can be divided into two subclasses: the Cyg-like sources
(\cyg2, \gx{5$-$1}, and \gx{340$+$0}) and the Sco-like sources
(\sco1, \sco2, and \gx{17$+$2}). Although the Z tracks of these
sources are similar, in CDs the horizontal branches of the Sco-like
sources are shorter and more inclined, while in HCDs their flaring
branches extend toward higher countrates. Also, no HBO has been
detected in \sco1 or \sco2, while the logarithmic derivative of the
centroid frequency with respect to countrate is \about 4 for the weak
HBO seen in \gx{17$+$2}, \about 3 times larger than in the Cyg-like
sources (see Hasinger \& van der Klis 1989).

Our spectral model qualitatively reproduces the spectral behavior of
the Sco-like Z sources if \emax\ is always \aboutless 0.5~keV in
these sources, rather than \about 0.6-1.0~keV as in the Cyg-like
sources, since the shape of the spectrum in the \exosat\ energy range
is then insensitive to changes in \emax. The increase in \taurf\ with
increasing \mdot\ on the horizontal branch then causes the branch to
slope downward to the right in CDs, while the increase of \tehcc\ and
\taurf\ on the flaring branch causes the countrate in the \exosat
energy range to increase, creating a flaring branch that extends to
higher countrates than in the Cyg-like Z sources.

The lower value of \emax\ needed to reproduce the Z tracks of the
Sco-like sources implies that their magnetic fields are smaller than
the magnetic fields of the Cyg-like sources. This is consistent with
the fact that no HBO has been detected in \sco1 or \sco2 and that the
HBO observed in \gx{17$+$2} is very weak. Moreover, according to the
unified model the logarithmic derivative of the HBO centroid
frequency with countrate should be greater for neutron stars with
lower magnetic fields (Ghosh \& Lamb 1992), in agreement with the
behavior of the HBO frequency observed in \gx{17$+$2}.

Finally, we mention that the spectral model described here reproduces
the observed X-ray colors of the atoll sources if their magnetic
fields are as weak as those of the Sco-like Z sources or weaker, and
their mass accretion rates are \about 0.01--0.1\ts\mdote.

A more detailed account of our spectral model and its application to
the Cyg-like and Sco-like Z sources and other accreting neutron stars
in low-mass systems will be presented in a forthcoming paper.

\acknowledgements

It is a pleasure to thank Gordon Baym, Vicky Kalogera, and Michiel van
der Klis for useful discussions. We also thank Michiel van der Klis
for providing \exosat data on \cyg2. This work was supported in part
by NSF grant AST~93-15133 and NASA grants NAGW 1583, NAG 5-2925, and
NAGW 2935.

\newpage
\figcaption[]{Solid lines: Values of the spectral parameters
\emax, \yhcc, and \taurf\ used in calculating the Z tracks shown in
Fig.~2. Dashed lines: Values of \emax\ and \yhcc\ computed assuming,
as an example, that the stellar magnetic field is 7\ee9~G, that the
radius of the magnetosphere as a function of \mdot\ is given by
disk-magnetosphere interaction model 2B of Ghosh \& Lamb (1992), and
that the electron density in the cyclotron-radiating region is
constant on the horizontal branch, increases from 10\sup{19} to
8\ee{19}~\pcc down the normal branch (due to the rapid reduction of
the inflow velocity caused by the increase in the outward radiation
force), and increases a further 2\% on the flaring branch. The
bottom panel shows the accretion rate inferred from the spectral
model and \exosat countrates, in units of the accretion rate
$\dot{M}_2$ at rank number 2; in the physical model, $\dot{M}_2 =
\mdote$.}

\figcaption[]{(a)~Comparison of the color-color track produced by
the spectral parameter variations shown in Fig.~1 with \cyg2 data
from \exosat. The colors are those used by Hasinger \& van der Klis
(1989). (b)~Similar comparison of hardness-countrate tracks. As usual
for \exosat data, the flaring branch lies almost on top of the normal
branch in the hardness-countrate diagram. (c)~Comparison of the
predicted countrate spectrum at rank number 1.5 with a time-averaged
\exosat countrate spectrum from near the middle of the normal
branch.}

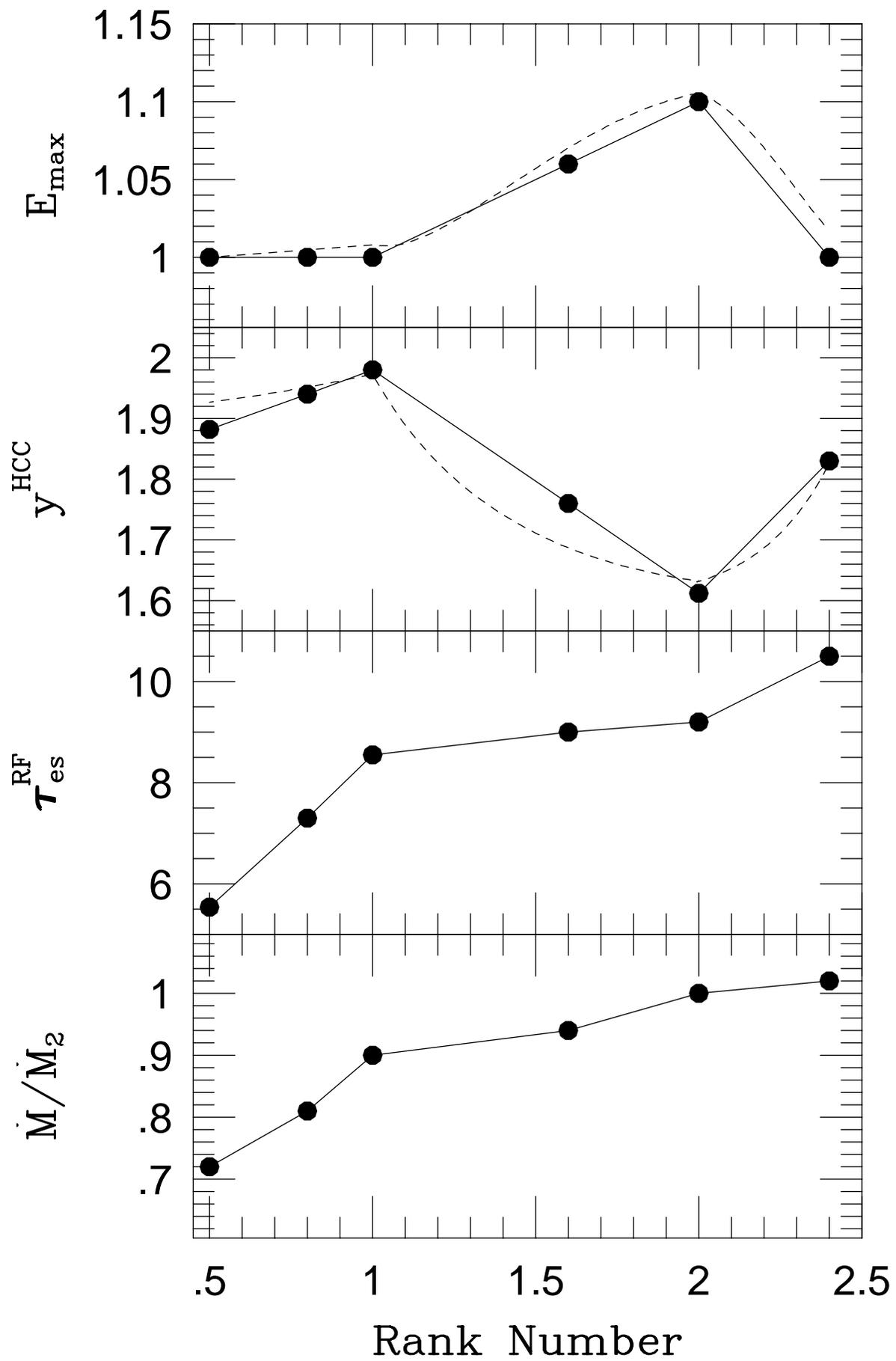
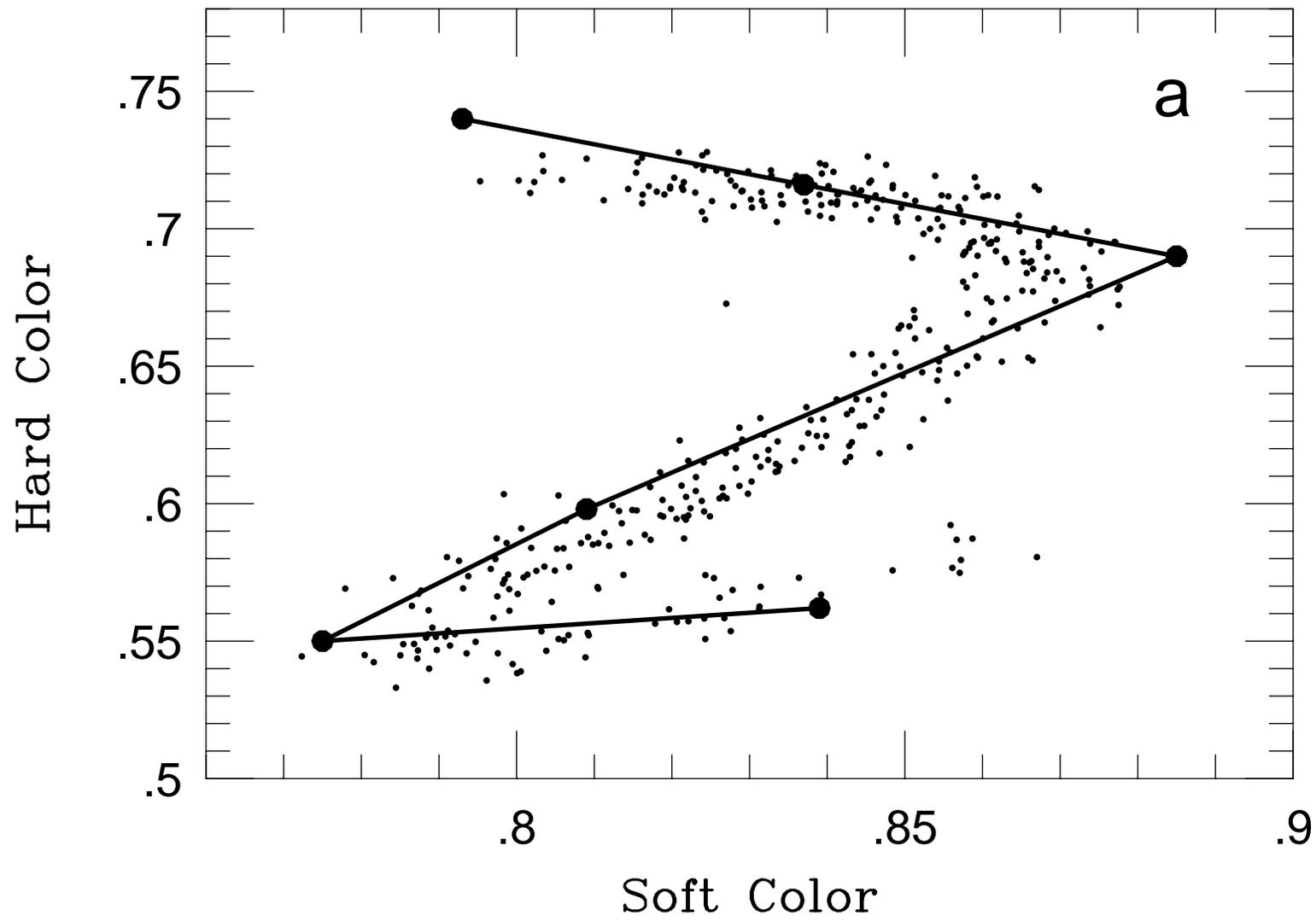
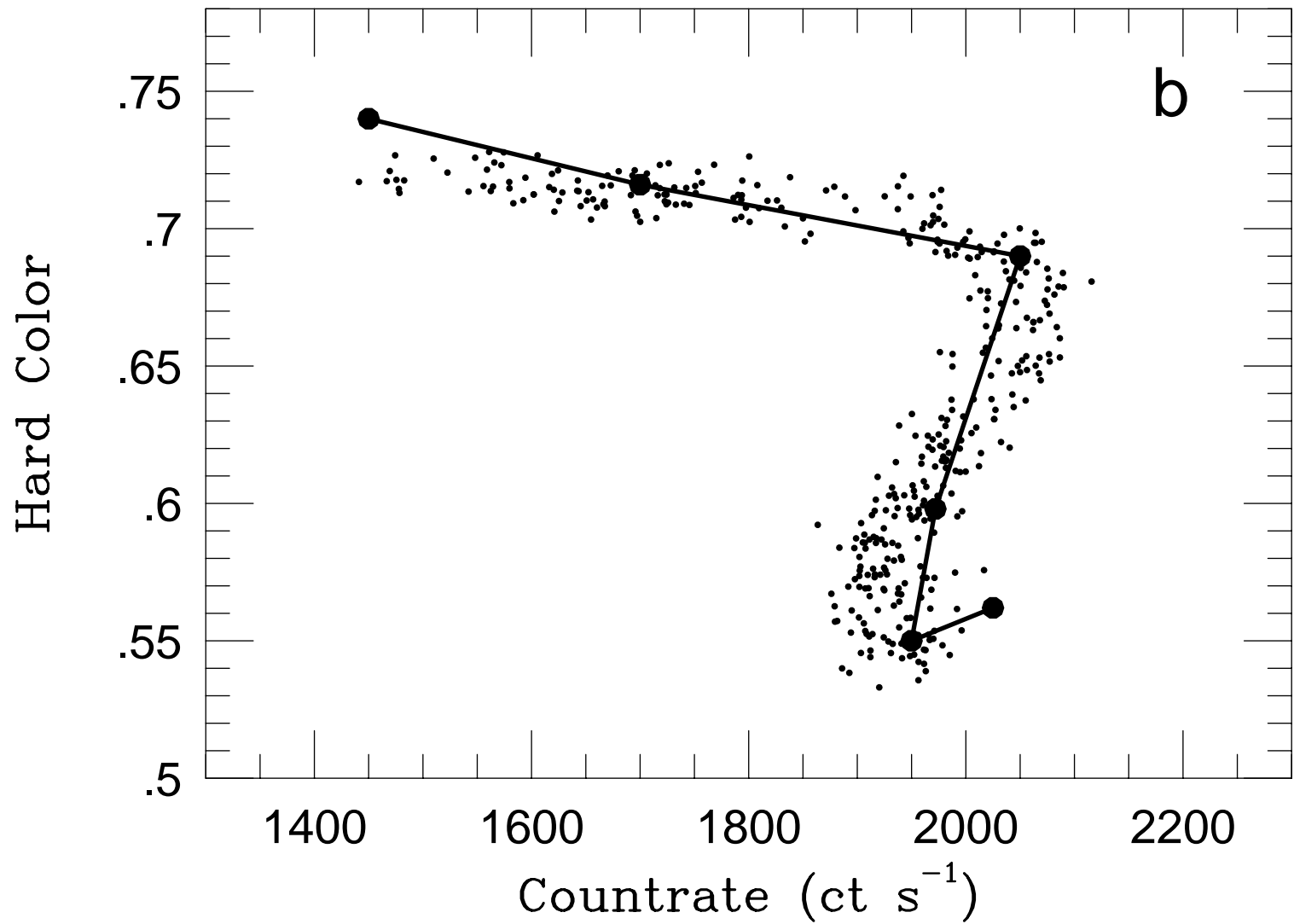
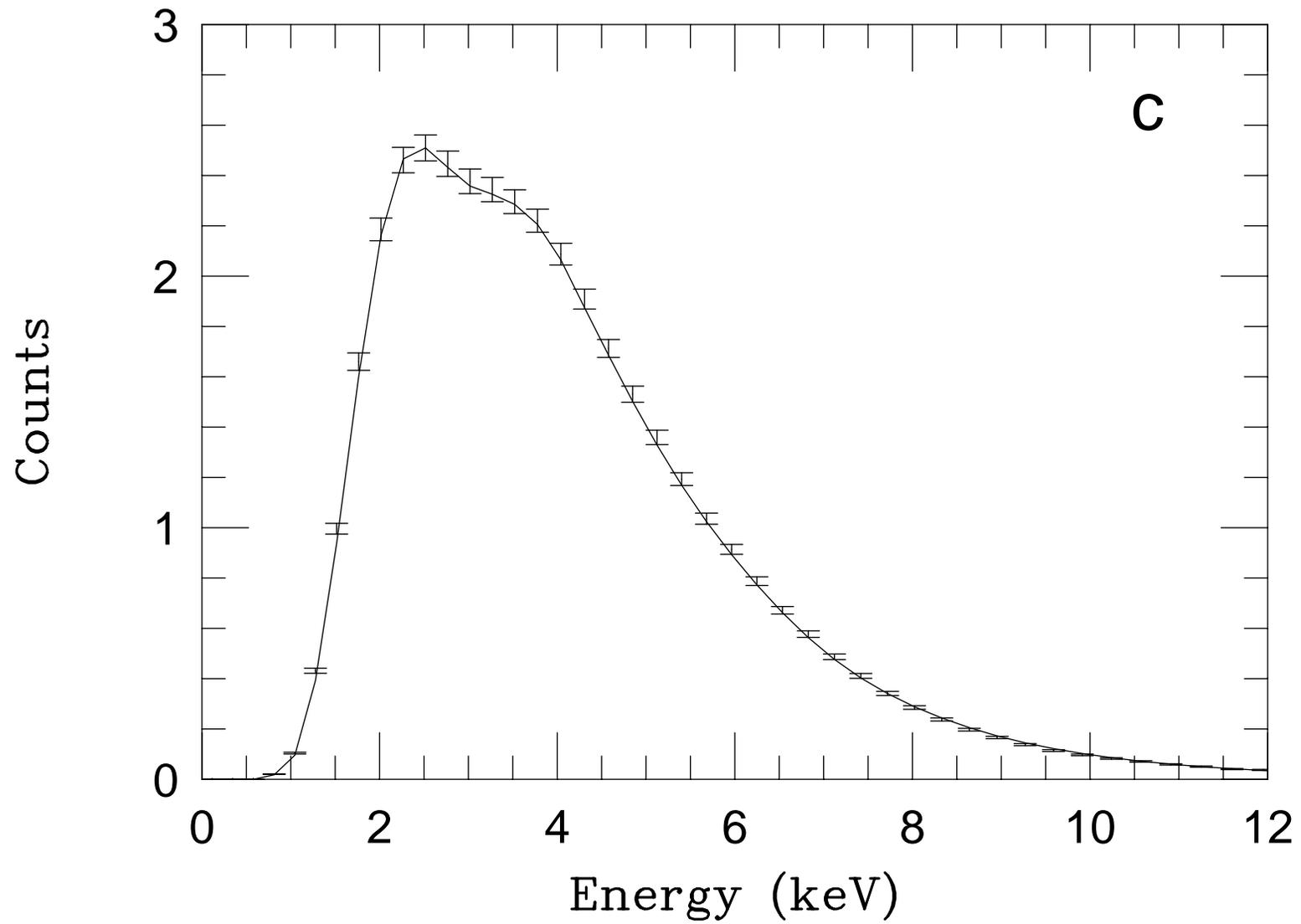
\end{document}